\documentclass{article}
\usepackage{jheppub}
\usepackage[utf8]{inputenc}
\usepackage[english]{babel}
\usepackage{amssymb}
\usepackage{mathtools}
\usepackage{graphicx}
\usepackage{float}
\usepackage{cancel}
\usepackage{enumitem}
\usepackage{appendix}
\usepackage{simplewick}
\usepackage{appendix}
\usepackage{bbm}
\usepackage{braket}
\usepackage{subfigure}

\usepackage{footmisc}
\usepackage{epsf,amsfonts,slashed,soul}
\usepackage{lipsum}
\usepackage{pifont}
\usepackage{bbold}
\usepackage{dsfont}
\usepackage{tikz}
\usepackage{verbatim}
\usepackage{placeins}

\newcommand{\Tr}{\text{Tr}}

\newcommand{\bea}{\begin{eqnarray}}
\newcommand{\eea}{\end{eqnarray}}

\newcommand{\abs}[1]{|#1|}

\newcommand{\nn}{\nonumber}

\newcommand{\be}{\begin{equation}}
\newcommand{\ee}{\end{equation}}
\newcommand{\Op}{\mathcal{O}}
\newcommand{\lv}{\mathcal{L}}

\title{Probing the Entanglement of Operator Growth}
\author[a]{Dimitrios Patramanis}
\emailAdd{d.patramanis@uw.edu.pl}
\affiliation[a]{Faculty of Physics, University of Warsaw, ul. Pasteura 5, 02-093 Warsaw, Poland.}

\abstract{
    In this work we probe the operator growth for systems with Lie symmetry using tools from quantum information. Namely, we investigate the Krylov complexity, entanglement negativity, von Neumann entropy and capacity of entanglement for systems with SU(1,1) and SU(2) symmetry. Our main tools are two-mode coherent states, whose properties allow us to study the operator growth and its entanglement structure for any system in a discrete series representation of the groups under consideration. Our results verify that the quantities of interest exhibit certain universal features in agreement with the universal operator growth hypothesis. Moreover, we illustrate the utility of this approach relying on symmetry as it significantly facilitates the calculation of quantities probing operator growth. In particular, we argue that the use of the Lanczos algorithm, which has been the most important tool in the study of operator growth so far, can be circumvented and all the essential information can be extracted directly from symmetry arguments.}

\begin{document}
\maketitle

\section{Introduction}
 Quantum information tools are becoming an increasingly important resource for the study of complex systems. In fact, quantities from quantum information theory have found applications in a variety of topics ranging from condensed matter physics to the study of quantum gravity. Among those, one of the most well-studied is entanglement entropy which has been at the center of many pioneering works. For example, in the context of the AdS/CFT correspospondence \cite{Maldacena:1997re}, the seminal works of Ryu and Takayanagi \cite{Ryu2006AspectsEntropy,Ryu2006HolographicCorrespondence} have established the duality between the entanglement entropy of intervals in the CFT and the length of bulk geodesics. This development has sparked many subsequent works further exploring the relationship between entanglement entropy and the structure of spacetime \cite{vanRaamsdonk2010BuildingEntanglement,Lewkowycz2013GeneralizedEntropy,Hubeny2007AProposal,Maldacena2013, Faulkner2013QuantumEntropy,Czech:2016xec,Chen:2021lnq,Rangamani:2016dms}. Considerations of entanglement entropy for many-body and condensed matter systems have been equally important in furthering our ability to understand them and predict their behavior, as reviewed in \cite{Amico:2007ag}. 
 
 In recent years this discourse has grown especially in the direction of systems with chaotic dynamics. More specifically, what one finds in such cases is that entanglement entropy might not be sufficient for a complete description of the system. This observation holds particular relevance in the study of black holes, where it appears that more refined probes are required to describe their late time dynamics. In \cite{Susskind:2014moa}, it was suggested that a quantity which could serve this purpose is computational complexity, which appears to be correctly capturing the features of the black hole evolution at late times. Since then significant effort has been devoted to finding appropriate measures of complexity and studying their properties \cite{Brown:2015bva,Brown:2015lvg,Jefferson:2017sdb,Chapman:2017rqy,Caputa:2017yrh,Caputa:2018kdj,Chapman:2021jbh, Balasubramanian:2019wgd,Chen:2020nlj,Chagnet:2021uvi,Abraham1987FoundationsMechanics,Erdmenger:2020sup}. 
 
 The main topic of this work is the study of operator growth, which has been investigated from different angles in a number of recent works \cite{Parker2019AHypothesis,Roberts:2018mnp,Dymarsky2020QuantumSpace,Dymarsky:2021bjq,Rabinovici2021OperatorSpace,Kar:2021nbm,Caputa:2021sib,Caputa:2021ori,Jian:2020qpp,Magan:2020iac,Barbon:2019wsy,MacCormack:2020auw,Kim:2021okd,Carrega:2020jrk}. As the name suggests, operators in quantum systems tend to ``grow" in the sense that they become more complicated as the system evolves in time, even if we start with some operator that is initially simple. This is particularly prominent in chaotic systems for which this growth is believed to be maximal. We will make these statements more precise in following sections but the question we ought to answer from the outset is in what way can one probe this type of process. We advocate that certain tools from quantum information are exceptionally well-suited candidates as they highlight universal features of operator growth. Additionally, we show that computing these probes can be greatly facilitated for systems with symmetry. More concretely, our goal will be to characterize operator growth for systems with SU(1,1) and SU(2) symmetry. We will achieve this by employing the generalized coherent states associated to each group, which will allow us to compute their Krylov complexity, entanglement entropy, negativity, and capacity of entanglement. The main point of comparison between these two cases will be the chaotic and integrable dynamics that they exhibit respectively. 
 
 To this end, we have structured the article as follows. In section 2 we provide a brief review on the topic of operator growth and how it can be characterized using Krylov complexity. Section 3 reviews the construction of two-mode coherent states which serve as the starting point for the calculations that we need to carry out. Section 4 contains a brief introduction to the other quantum information tools mentioned above and our main results for SU(1,1) and SU(2). Finally, in section 5 we summarize our conclusions and provide additional comments.

\section{Operator growth and Krylov complexity}
Studying the time evolution of operators in quantum systems is hardly a new problem. However, recently there has been a renewed interest in the subject mainly in the context of chaotic systems. One of the developments that has been highly influential is the universal operator growth hypothesis \cite{Parker2019AHypothesis}, which asserts that there are certain features of operator growth that depend only on the class of model under study and not on its specific details. In this section we will briefly review the notion of operator growth in general and follow the arguments leading up to the universal operator growth hypothesis. A quantity that arises naturally in this endeavor is Krylov complexity (sometimes referred to as K-complexity), which will also prove relevant to our discussions in later sections by virtue of it being a quantum information tool of particular interest. There are a number of works with pedagogical introductions to this topic \cite{Parker2019AHypothesis,Caputa:2021sib,Rabinovici2021OperatorSpace,Dymarsky2020QuantumSpace}, so here we restrict ourselves to the basic notions required for the ensuing discussions.

Let us start by stating the problem that we wish to study. Suppose we are given a quantum system described by the Hamiltonian $H$ and a time-dependent operator $\Op (t)$. For simplicity we will assume that the Hamiltonian is local and the operator Hermitian. More general setups have been considered for example in \cite{Dymarsky2020QuantumSpace}, however for the purposes of the present work it suffices to restrict to the above. We shall now try to determine the time evolution of the operator $\Op (t)$ which in the Heisenberg picture is given by
\be
\Op (t) = e^{iHt}\Op (0)e^{-iHt}~,
\ee
or equivalently as an expansion 
\be 
\Op (t)= 1+it[H,\Op(0)]+\frac{(it)^2}{2}[H,[H,\Op(0)]]+...
\ee
We would like to have an expression for $\Op (t)$ at arbitrary times and obviously having to perform all the nested commutators is not the optimal way of doing so. A particular way to approach this problem is by using the Lanczos algorithm \cite{Lanczos1950AnIM,Viswanath1994TheRM}. This is simply an iterative procedure that using some initial data can generate for us approximate answers. The main idea behind the Lanczos algorithm is providing an orthonormal basis in terms of the nested commutators of the Hamiltonian, which we can then use to express $\Op (t)$ at an arbitrary time. To make this more precise let us first define the Liouvillian super-operator
\be
\lv= [H,\cdot]~.
\ee
This then implies
\be \label{ liouvillian expansion}
\Op (t)=e^{i\lv t} \Op (0)= 1+ it\lv \Op (0) +\frac{(it)^2}{2}\lv^2 \Op (0)+ ...
\ee
We seek to construct a basis out of the elements 
\be
\Op(0)=|\Tilde{\Op}), \quad \lv \Op (0)= |\Tilde{\Op}_1),\quad \lv^2 \Op (0) =|\Tilde{\Op}_2)...
\ee
To do so it is convenient to view the above as the basis vectors (states) in an abstract operator Hilbert space provided that we can orthogonalize them with respect to each other. We denote the orthonormalized vectors as $|\Op_n)$. We must also specify the inner product between these vectors. The standard choice used throughout the literature is the Wightman inner product 
\be
(A|B)=\langle e^{ H\beta/2}A^\dagger e^{-H\beta/2}B \rangle_\beta~,
\ee
where $\langle\rangle_\beta$ is to be understood as the thermal expectation value at inverse temperature $\beta$. Having laid down the groundwork, one only needs to provide a scheme for the orthogonalization of the vectors $|\Op _n)$ which is precisely the purpose of the Lanczos algorithm. Starting with the initial state $|\Op)$ we subsequently obtain 
\be
|\Op_1)=b_1^{-1}\lv|\Op)~,
\ee
where $b_1$ is the normalization constant. The algorithm proceeds iteratively as follows
\be
|A_n)=\mathcal{L}|\Op_{n-1})-b_{n-1}|\Op_{n-2})~,\label{AnDef}
\ee 
and then normalizing
\be
|\Op_n)=b^{-1}_n|A_n),\qquad b_n=(A_n|A_n)^{1/2}~.
\ee

In this manner one orthogonalizes each vector with respect to the previous one in the sequence and ultimately obtains an orthonormal basis, referred to as the Krylov basis. The normalization constants $b_n$ are called the Lanczos coefficients and as we will see below they encode very useful information about the behavior of the system. Depending on the system of interest the Lanczos algorithm may or may not terminate, providing a finite or infinite dimensional Krylov space respectively. Once this procedure has been completed the operator $\Op (t)$ can be expanded using the Krylov basis as 
\be
|\Op(t))=\sum_n i^n \varphi_n(t)|\Op_n)\label{OTKrB}~,
\ee
where $ \varphi_n(t)$ are some appropriate coefficients that satisfy the condition 
\be
\sum_n |\varphi_n(t)|^2=1~.
\ee
Of course determining these coefficients is not a menial task. However, with a few lines of algebra one can find a formula that relates them to the Lanczos coefficients (for a detailed derivation see \cite{Caputa:2021sib}). This expression takes the form of a discrete Schrodinger equation that is solved iteratively 
\be\label{SchrodingerEq}
\partial_t\varphi_n(t)=b_n\varphi_{n-1}(t)-b_{n+1}\varphi_{n+1}(t)~.
\ee

Already from (\ref{ liouvillian expansion}) it is easy to see that as the system evolves in time we require more terms of the expansion to accurately describe the operator. From the perspective of the Krylov basis this can be interpreted as the operator ``growing" in Krylov space and as such requiring more basis vectors for its decomposition. A natural candidate that can serve as a quantitative measure of this growth is Krylov complexity, simply defined as 
\be
K_{\mathcal{O}}\equiv\sum_{n} n \,\vert \varphi_n(t)\vert^2~.
\ee
Equation (\ref{SchrodingerEq}) can be thought of as describing a particle moving on a one-dimensional chain, where $\varphi_n$ are the different sites and $b_n$ are the hopping coefficients between them. This entails an elegant interpretation of Krylov complexity as the average position of the particle on the chain. 

Before concluding this section a few remarks are in order. We have described a procedure that enables us to extract information about the evolution of Heisenberg operators in principle for any quantum system. More importantly it is possible to take an extra step and classify quantum systems according to certain universal features that they exhibit. Namely, for chaotic systems one expects that the growth of operators will be maximal, whereas for other generic cases such as integrable or free theories the growth will proceed at a slower rate. Hopefully it will become clearer why that should be the case in later sections, but for now let us rely on our physical intuition to justify these claims. As we mentioned previously the problem of operator growth can be mapped to a particle hopping on a one-dimensional chain. From this point of view one expects that it becomes increasingly easier for the particle to hop on every next site, since that would be the picture consistent with an operator that exhibits maximal growth. Hence, the Lanczos coefficients $b_n$ that serve as the hopping amplitudes should grow with increasing $n$. The question is then ``how fast do they grow"? The simple answer provided by the universal operator growth hypothesis for maximally chaotic theories is ``as fast as locality permits". More concretely, in \cite{Parker2019AHypothesis} the authors show that, following this hypothesis, the growth of $b_n$ is of the linear form $b_n\sim an+\gamma$, which in turn implies an exponential growth of Krylov complexity with a characteristic exponent that depends on the specifics of the system. This result is consistent with the notion of classical chaos where one also expects an exponential growth with a characteristic Lyapunov exponent. Similar conclusions can be drawn for integrable and free theories where the $b_n$ have been shown to grow as $b_n\sim \sqrt{n}$ and remain constant respectively. More recently, the authors of \cite{Dymarsky:2021bjq} have shown that certain free field QFTs attain the maximal growth for the Lanczos coefficients and hence they have argued that this should not be necessarily regarded as a sign of chaos. While this shows that the universal operator growth hypothesis cannot immediately discriminate between chaotic and non-chaotic theories, it still provides a framework within which one can have a universal description of different classes of systems according to the time evolution of their operators.  

However, such powerful techniques always come with certain limitations. In this case we are restricted by the iterative nature of this approach. Even though it is very well-tailored for obtaining numerical results, it is generally hard to go beyond that. In a handful of cases it is possible to obtain a closed form for the Lanczos coefficients, but even for those the process is far from simple.

Recently, this obstacle was partially circumvented for systems whose symmetry is described by a Lie group. More specifically, in \cite{Caputa:2021sib} it was shown that it is possible to obtain the Lanczos coefficients as well as the $\varphi_n$ and quantities related to those directly from symmetry arguments without appeal to the Lanczos algorithm itself. The way to do so is through coherent states, which are objects intimately connected to the symmetry of the problem. In the sections to come we will further explore this direction and show how we can use it to not only compute the Krylov complexity, but other quantum information tools as well. 

\section{Two-mode coherent states}

In this section we review the construction of two-mode coherent states in the context of the Lie groups that are examined in this work. This will allow us to compute several quantities of interest in the following sections. The key fact in this discussion is that for each group under consideration there exist discrete series representations that can be obtained by expressing the generators of the group in terms of a pair of bosonic ladder operators. Let us make this assertion more concrete by first examining the properties of SU(1,1) (a more rigorous discussion can be found in \cite{Perelomov:1986tf}). 
\par

The Lie algebra of SU(1,1) has three generators $K_0,K_1,K_2$ satisfying the commutation relations 
\be
[K_1,K_2]=-iK_0, \quad [K_2,K_0]=iK_1, \quad [K_0,K_1]=iK_2~.
\ee
We can define a set of ladder operators by simply changing the basis as follows
\be
K_\pm=\pm i(K_1 \pm i K_2), \quad K_0~.
\ee
The appropriate basis vector on which these operators act is of the form $\ket{k,\mu}$. For the discrete series representations $k$ takes integer and half-integer values ($k=1/2,1,3/2,...$) and $\mu$ is the eigenvalue of $K_0$. We can further modify this basis by the identifications
\begin{equation}
  K_+=a^\dagger b^\dagger,\quad K_-=a b,\quad K_0=\frac{1}{2}(a^\dagger a + b^\dagger b +1)~,   
\end{equation}
where $a,b$ are bosonic ladder operators that satisfy the usual commutation relations $[a,a^\dagger]=[b,b^\dagger]=1$. It is straightforward to check that this new basis still satisfies the SU(1,1) commutation relations and naturally the associated basis vector has the form of a two-mode state
\be
\ket{m,n}=(m!n!)^{-\frac{1}{2}}(a^\dagger)^m(b^\dagger)^n\ket{0,0}~.
\ee
The characteristic $k$ of the representation is related to the above expression through $k=\frac{1}{2}(1+|{n_0}|)$, where $n_0=m-n$. Therefore, it is possible to obtain any representation of the discrete series by considering the appropriate two-mode state. 

Using the Fock space that each representation defines one can construct an associated family of coherent states. This is achieved by the action of a displacement operator on the appropriate vacuum state. The former is defined in terms of the algebra generators as:
\be
D(\xi)=e^{\xi K_+-\Bar{\xi}K_-}=e^{\xi a^\dagger b^\dagger-\Bar{\xi}ab}=e^{zK_+}e^{\eta K_0}e^{-\Bar{z}K_-}~,
\ee
where in the last equality we used the BCH formula to bring the operator in a normal form with
\be
z=\frac{\xi}{\abs{\xi}}\tanh{\abs{\xi}}, \quad \eta= 2\ln{\cosh{\abs{\xi}}} ~.
\ee
It is useful to introduce polar coordinates $\xi=re^{i\phi}$, such that $z$ parametrizes the unit disc
\be
z=e^{i\phi}\tanh{r} , \qquad \abs{z}<1 ~.
\ee
The coherent states that we are about to construct are actually well studied objects in the field of quantum optics (for a pedagogical introduction see \cite{Agarwal2012QuantumOptics}). They are referred to as two-mode photon added (or subtracted) squeezed states and despite their long name are actually rather simple. The most common species among them are the single photon added states which arise from the action of the above displacement (squeezing) operator on the shifted vacuum $\ket{1,0}$. Here we want to consider their more exotic cousins that are generated from an arbitrarily shifted vacuum $\ket{n_0,0}$. Thus,  the coherent states we are seeking are given by 
\be
\ket{z}_{n_0}=\mathcal{N}_{n_0} (b)^{n_0} D(z)\ket{0,0}~,
\ee
where $\mathcal{N}$ is a normalization constant. We would like to express $\ket{z}$ as a linear combination of Fock states, for which we need the Bogoliubov transformation

\be
b(z)=D^\dagger(z)b D(z)=b\cosh(r)+a^\dagger e^{i\phi}\sinh(r)~.
\ee
Using the above it is straightforward to obtain
\be
 \ket{z}_{n_0}=\mathcal{N}_{n_0} D(z)e^{i{n_0} \phi}\sinh^{n_0}{r}\sqrt{n_0!}\ket{{n_0},0}~.
\ee
The normalization is chosen as
\bea
\mathcal{N}_{n_0}&=&\bra{z}(b^\dagger)^{n_0}(b)^{n_0}\ket{z}^{-1/2}=\bra{0,0}D^\dagger(z)(b^\dagger)^{n_0}(b)^{n_0} D(z)\ket{0,0}^{-1/2},\nn\\
&=&\bra{0,0}(b^\dagger(z))^{n_0} (b(z))^{n_0}\ket{0,0}^{-1/2}
\eea
and an explicit computation yields
\be
\mathcal{N}_{n_0}=\frac{1}{\sqrt{n_0!}\sinh^{n_0}{r}}~.
\ee
We can now act with the displacement operator on the shifted vacuum to get
\begin{align}
    D(z)e^{i {n_0} \phi}\ket{{n_0},0}&=e^{i{n_0} \phi}e^{e^{i\phi}\tanh(r)a^\dagger b^\dagger}e^{-\log(\cosh(r))(a^\dagger a+b^\dagger b+1)}e^{-e^{i\phi}\tanh(r)ab}\ket{{n_0},0}\\
    &=e^{i{n_0} \phi}\frac{1}{(\cosh{r})^{{n_0} +1}}\sum_{n=0}^\infty e^{i n\phi}(\tanh{r})^n\sqrt{\binom{n+n_0}{n_0}}\ket{{n_0}+n,n}~.
\end{align}
 Therefore, neglecting the constant phase factor $e^{in_0\phi}$, which as we shall see is irrelevant for our subsequent calculations, we can write
\be \label{sl2 states}
\ket{z}_{n_0}=\sum_{n=0}^\infty e^{in\phi}\varphi_n\ket{n_0+n,n}~,
\ee
with
\be
\varphi_n=\frac{\tanh^n{r}}{(\cosh{r})^{{n_0} +1}}\sqrt{\binom{n+n_0}{n_0}}~.
\ee
The choice of the symbol $\varphi$ to represent the coefficients of the decomposition is of course not a coincidence. In \cite{Caputa:2021sib} it is rigorously shown that these coefficients match precisely the wavefunctions $\varphi$ that arise from the Lanczos algorithm as was explained in the previous section. Furthermore, the variable $r$ is taken to be proportional to time $r=\alpha t$, which allows an interpretation of the operator growth in terms of a motion in the classical phase space of the problem. In this work we will also treat $r$ as time even though we will not denote it explicitly. For an alternative derivation of these states using the techniques of 2d CFT see appendix \ref{appendix}.  

Let us now turn our attention to the construction of the two-mode coherent states for SU(2) \cite{Mathur2001CoherentSU3}. The process is practically the same as before, so for the sake of brevity we will skip the tedious steps and focus on the essentials. The Lie algebra of SU(2) is characterized by 3 generators that we label $J_0,J_+,J_-$ satisfying the commutation relations
\begin{equation}
    [J_0,J_\pm]=\pm J_\pm ,\qquad [J_+,J_-]=2J_0~.
\end{equation}
Once again we seek to express those in terms of a pair of bosonic ladder operators which we will now label $\alpha_1,\alpha_2$ to keep them distinct from the SU(1,1) case. To build a representation of SU(2) the appropriate relation between the bosonic operators and the $J_0,J_+,J_-$ is given by
the following identifications 
\be
J_+=\alpha^\dagger_1 \alpha_2, \quad J_-=\alpha^\dagger_2 \alpha_1,\quad J_0=\frac{1}{2}(\alpha^\dagger_1 \alpha_1+\alpha^\dagger_2 \alpha_2)~.
\ee
The coherent states in this basis are given by
\begin{equation}
    \ket{z_1,z_2}=\sum_{N_1,N_2}F_{N_1,N_2}\ket{N_1,N_2}~,
\end{equation}
where the function $F$ is
\be
F_{N_1,N_2}=\sqrt{\frac{(N_1+N_2)!}{N_1!N_2!}} (\cos{\chi}e^{i \beta_1})^{N_1}(\sin{\chi}e^{i \beta_2})^{N_2}~.
\ee
The coordinates $0\leq\chi\leq\pi/2, \quad 0\leq\beta_1\leq2\pi,\quad 0\leq\beta_2\leq2\pi$ parametrize the 3-sphere and the occupation numbers $N_1,N_2$ have to satisfy the property $N_1+N_2=2j$ where j is the quantum number associated to the typical representation of SU(2) mentioned above. Thus we can rewrite the expression for $F$ as:
\be
F_{N_2}=\sqrt{\frac{(2j)!}{(2j-N_2)!N_2!}} (\cos{\chi}e^{i \beta_1})^{2j-N_2}(\sin{\chi}e^{i \beta_2})^{N_2}~.
\ee
This will allow us to express our results in terms of the more familiar quantum number $j$. Notice that this implies that our states can be rewritten as 
\be 
 \ket{z_1,z_2}=\sum_{N_2=0}^{2j}F_{N_2}\ket{N_2,2j-N_2}~.
\ee
\section{Quantum information tools}
Before diving into the computations that utilize the technology developed above, we provide some generalities about each of the quantities that we will be computing in this section. Krylov complexity was reviewed in its own right in section 2, so here we will be concerned with negativity, entanglement entropy and capacity of entanglement. 

Negativity is a measure of entanglement that, given a mixed state $\rho$, quantifies by how much the partial transpose $\rho^{PT}$ fails to be positive definite \cite{Vidal2002ComputableEntanglement}. It is defined as 
\be
\mathcal{N} \equiv \frac{||\rho^{PT}||-1}{2}~,
\ee
where 
\be
||\rho^{PT}||=\Tr{\sqrt{{\rho^{PT}}^\dagger\rho^{PT}}}~.
\ee
A common variation (and the one we will be primarily concerned with) that originates from the above definition is the logarithmic negativity
\be
E_{\mathcal{N}}(\rho)=\log_2(1+2\mathcal{N}(\rho))~.
\ee
This particular probe of entanglement is widely used in quantum optics where it is common to know explicitly the density matrix. In other fields it is standard practice to use the natural logarithm instead of the logarithm with base 2, which simply results in a difference by some constant. 

For a general two mode state the computation is as follows. The density matrix is of the form
\be
\rho=\sum_{n,m}c_nc_m^*\ket{n,n}\bra{m,m}~.
\ee
The partial transpose is then given by transposing the elements of only one of the states obtaining
\be
\rho^{PT}=\sum_{n,m}c_nc_m^*\ket{n,m}\bra{m,n}~.
\ee
In \cite{Agarwal2012QuantumOptics} the negativity is computed by bringing the partial transpose to its diagonal form and reading off its eigenvalues. However, a straightforward calculation of the trace norm is more illuminating, albeit slightly longer. First, let us compute the Hermitian conjugate of $\rho^{PT}$
\begin{equation}
   {\rho^{PT}}^\dagger= \sum_{n,m} [c_nc_m^*\ket{n,m}\bra{m,n}]^\dagger= \sum_{n,m}c_n^*c_m\left[(\ket{n}\otimes\ket{m})(\bra{n}\otimes\bra{m})\right]^\dagger=\sum_{n,m}c_n^*c_m\ket{m,n}\bra{n,m}~.
\end{equation}
Computing the trace norm is then simply done as shown below
\begin{align}
 \abs{\abs{\rho^{PT}}}&= \sum_{k,l}\sqrt{\sum_{n,m}\abs{c_nc_m}^2\braket{k,l|m,n}\braket{n,m|n,m}\braket{m,n|l,k}}\\
 &=\sum_{k,l}\sqrt{\abs{c_kc_l}^2\braket{k,l|l,k}\braket{l,k|k,l}}\\
 &=1+\sum_{k\neq l}\abs{c_kc_l}
\end{align}
and therefore the logarithmic negativity assumes the simple form
\be
E_\mathcal{N}(\rho)=\log_2\left(1+\sum_{n\neq m}\abs{c_nc_m} \right) =\log_2\left(\sum_n\abs{c_n}\right)^2~.
\ee

Given a density matrix $\rho$ the entanglement entropy is defined as 
\be
S\equiv -\Tr\rho\ln\rho~,
\ee
however, in practice one usually obtains it as the $q\rightarrow 1$ limit of the Rényi entropy
\be
S^{(q)}=\frac{1}{1-q}\ln{\Tr\rho^q}~.
\ee

Finally, the capacity of entanglement is a concept recently gaining popularity in the context of the black hole information paradox \cite{Kawabata:2021hac,Kawabata:2021vyo}. This is because it is a probe that is more sensitive to the intricate phenomena that take place during black hole evaporation compared to the entanglement entropy which is the quantity that had been extensively studied in the past. An elaborate review on the capacity of entanglement was given in \cite{DeBoer:2018kvc}. There are several different definitions, but the one that is best suited for our purposes is given again in terms of the Rényi entropy as
\be
C=\frac{q^2d^2[(1-q) S^{(q)}]}{dq^2}\bigg|_{q=1}~.
\ee
In computing the Rényi entropies we will be taking advantage of the two-mode representation of our states. Namely, the density matrices of the states we are interested in are of the form $\rho=\abs{\varphi_n}^2\ket{n+n_0,n}\bra{n,n+n_0}$ and so by tracing over one of the two modes we have $\rho_n=\sum_n|\varphi_n|^2\ket{n+n_0}\bra{n+n_0}$. The $\abs{\varphi_n}^2$ can then be interpreted as the eigenvalues of the reduced density matrix, which in turn allows us to directly compute the Rényi entropies by the following substitution 
\be
S^{(q)}=\frac{1}{1-q}\ln{\Tr\rho^q}=\frac{1}{1-q}\ln{\sum_n \abs{\varphi_n}^{2q}}~.
\ee

Notice that all the quantities we are concerned with require as input the norm $\abs{\varphi_n}$, which implies that any phase factors in our expressions are rendered irrelevant. This is the reason we are being cavalier about keeping track of phase factors throughout this work, although it would be interesting to consider in the future whether they contain any non-trivial information.

\subsection{SU(1,1)}

\begin{figure}[ht]
    \centering
    \includegraphics[scale=0.6]{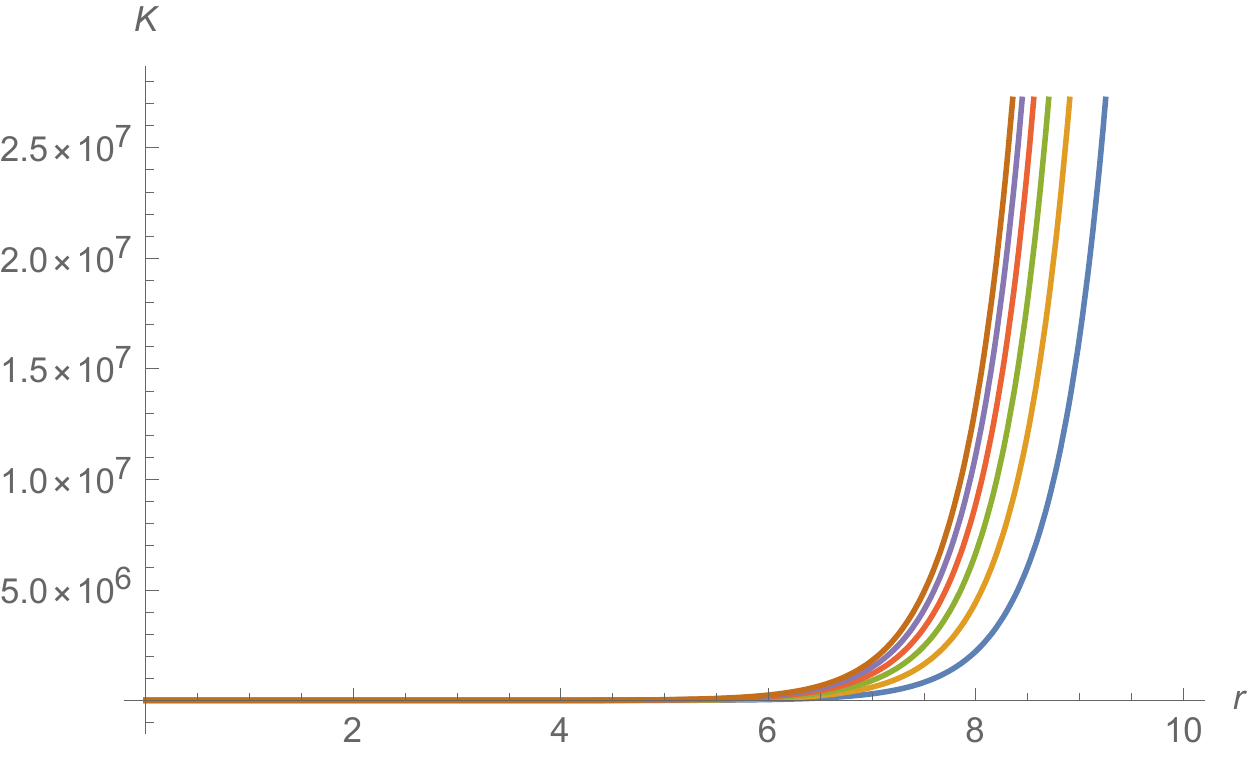}
    \caption{Krylov complexity for systems with SU(1,1) symmetry as a function of $r$. The curves depicted are for $n_0=0,1,2,3,4,5$. We observe that for bigger $n_0$ the exponential rise of the complexity starts sooner.}
    \label{krylov sl2}
\end{figure}

We begin with the easiest quantity to compute which is the Krylov complexity. By its definition, we simply have to perform the sum
\be
K=\sum_{n=0}^\infty n\abs{\varphi_n}^2=\sum_{n=0}^\infty\frac{(\tanh{r})^{2n}}{(\cosh{r})^{2({n_0} +1)}}\binom{n+n_0}{n_0}=(1+n_0)\sinh^2{r}~.
\ee
This clearly shows that for large $r$ (late times) the Krylov complexity grows exponentially as expected. This is depicted for a few different choices of $n_0$ in figure \ref{krylov sl2}.

\begin{figure}[ht]
    \centering
    \includegraphics[scale=0.6]{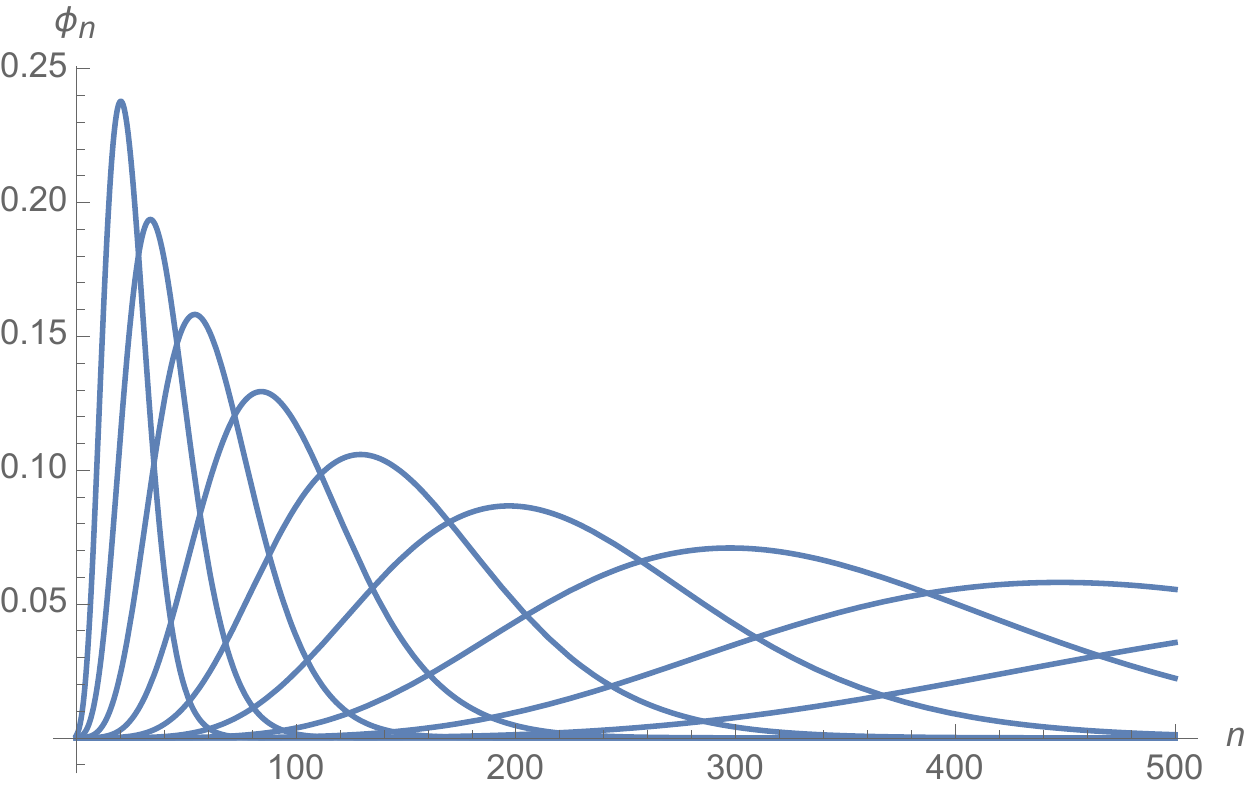}
    \caption{$\varphi_n$ as function of $n$. The chosen parameters for this graph are $n_0=2$ and $r$ takes values from 1 to 2.6 in steps of 0.2. This is essentially the spreading of the wavefunction that was observed in earlier works. }
    \label{r-dependence}
\end{figure}

\begin{figure}[ht]
\centering
\begin{subfigure}
\centering
    \includegraphics[width=0.45\textwidth]{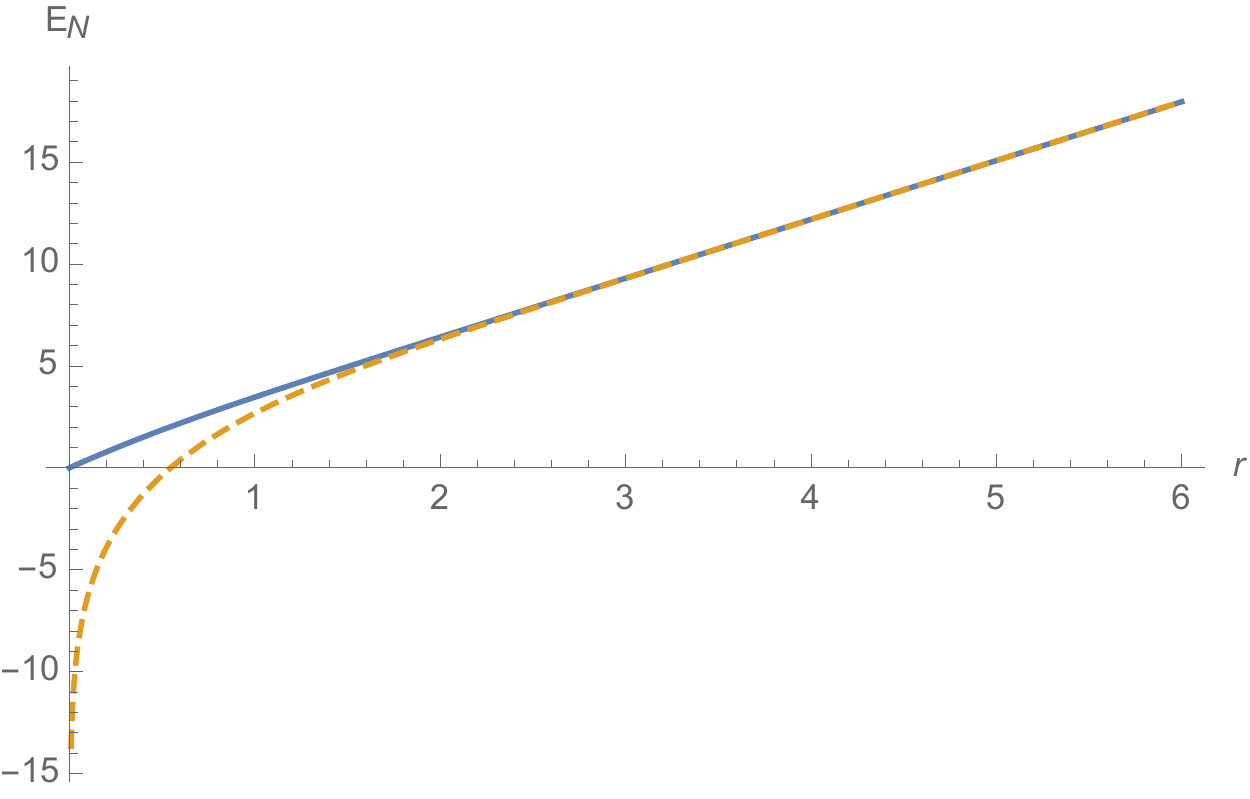}
\end{subfigure}
\begin{subfigure}
    \centering
    \includegraphics[width=0.45\textwidth]{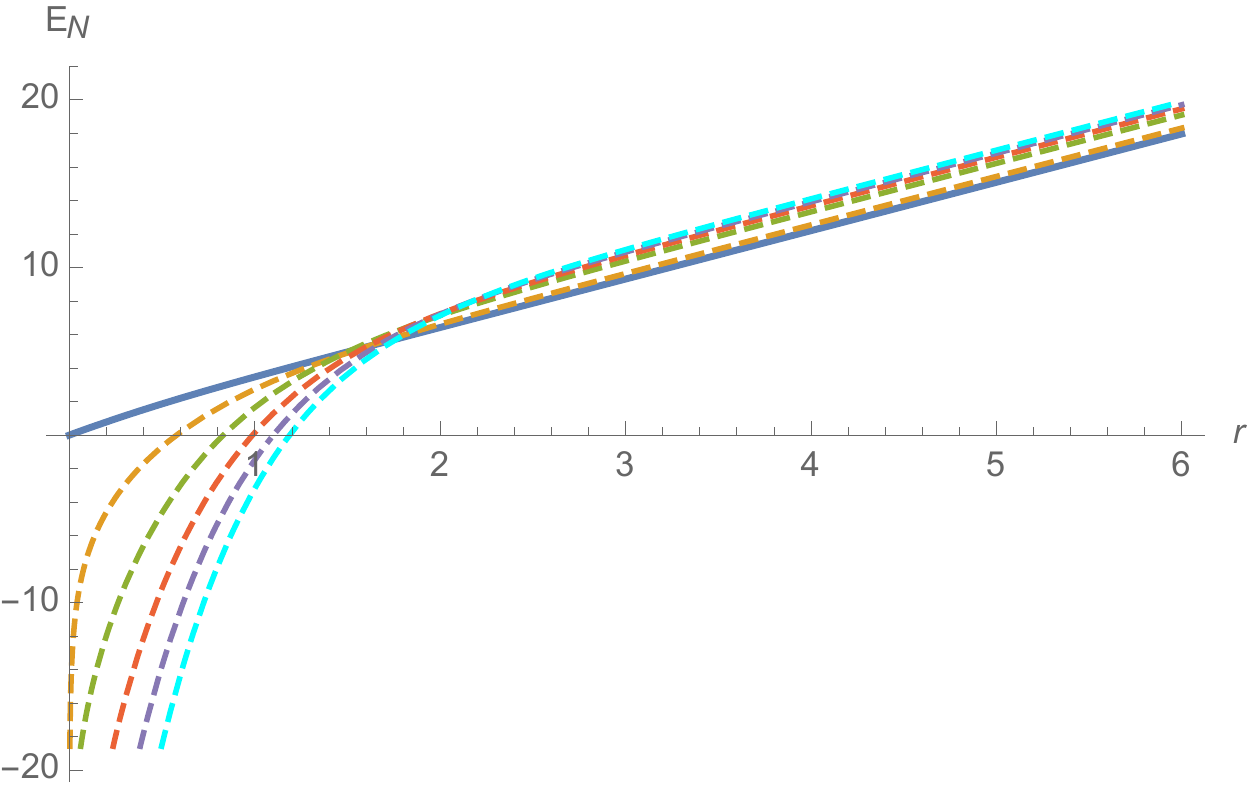}
\end{subfigure}
     \caption{Left: Comparison of the logarithmic negativity as a function of $r$ for $n_0=1$ between the exact solution (continuous curve) and the approximated solution (dashed curve). We maintain this way of representing exact results using continuous curves and approximate using dashed throughout the rest of this work. Right: Logarithmic negativity as a function of $r$ for $n_0=1,2,7,12,17,22$. Clearly, the features of the approximated functions cannot be trusted for early times as they become negative. However, for late times when our approximation becomes relevant, we observe a linear trend both for the exact and approximated solutions.}
     \label{sl2 neg}
\end{figure}

We proceed by computing the logarithmic negativity associated to the two-mode coherent states (\ref{sl2 states}). We showed that in general the former is given by 
\be
E_\mathcal{N}(\rho) =\log_2\left(\sum_n\abs{\varphi_n}\right)^2~,
\ee
and substituting the coefficients $\varphi_n$ it can be written out explicitly as 
\be
E_\mathcal{N}(\rho)=\log_2{\left(\sum^\infty_{n=0}\frac{\tanh^n{r}}{\cosh^{n_0+1}{r}}\sqrt{\binom{n+n_0}{n_0}}\right)^2}~.
\ee
For general $n_0$ an analytic solution does not appear to be possible and for that reason we will have to perform an approximation based on the behaviour of the $\varphi_n$ as functions of $r$. More concretely, we will show that for sufficiently large $r$ the functions have support in large enough $n$ to justify keeping only the leading order contribution of the binomial coefficient. In figure \ref{r-dependence} it is shown that as $r$ increases the relevant $n$ also increase. 
Therefore, one can expand the binomial coefficient for large $n$, which yields
\be
\binom{n+n_0}{n_0}\simeq\frac{n^{n_0}}{n_0\Gamma(n_0)}~.
\ee
For $n_0=0,1$ it is possible to obtain the analytic answers so let us compare them with the results of this approximation. For $n_0=0$ the approximation is exact so the comparison is trivial, but for $n_0=1$ we find out that the agreement for large $r$ is remarkable (as shown in figure \ref{sl2 neg}), thus justifying our initial arguments for using the large $n$ approximation.  
The solution for the negativity that we obtain for arbitrary $n_0$ using this approximation is 
\be \label{neg approx}
\log_2{\left(\sum^\infty_{n=0}\frac{\tanh^n{r}}{\cosh^{n_0+1}{r}}\sqrt{\binom{n+n_0}{n_0}}\right)^2}\simeq \log_2{\left(\frac{\text{Li}_{-\frac{n_0}{2}}(\tanh{r})}{\cosh{(r)}^{(1+n_0)}\sqrt{n_0\Gamma(n_0)}}\right)^2}~.
\ee
Since we are interested in the regime of large $n$, by virtue of our approximation, we can employ an asymptotic expansion of the negativity to obtain a more palatable expression in the case that $n_0$ is even. The answer we obtain is 
\be
n_0=2p\Rightarrow  E_\mathcal{N}=\log_2\left({\frac{4^{1-p}}{((\frac{3}{2})_{p-1})^2}}e^{2r}\right)~,
\ee
which makes more apparent that the negativity at late times follows a linear trend. In figure \ref{sl2 neg} one can see the logarithmic negativity resulting from (\ref{neg approx}) for different values of $n_0$. It is evident that these different cases exhibit a universal linear behaviour for large $r$, which is consistent with our expectations from the universal operator growth hypothesis.

\begin{figure}[ht]
    \centering
    \begin{subfigure}
    \centering
    \includegraphics[width=0.45\textwidth]{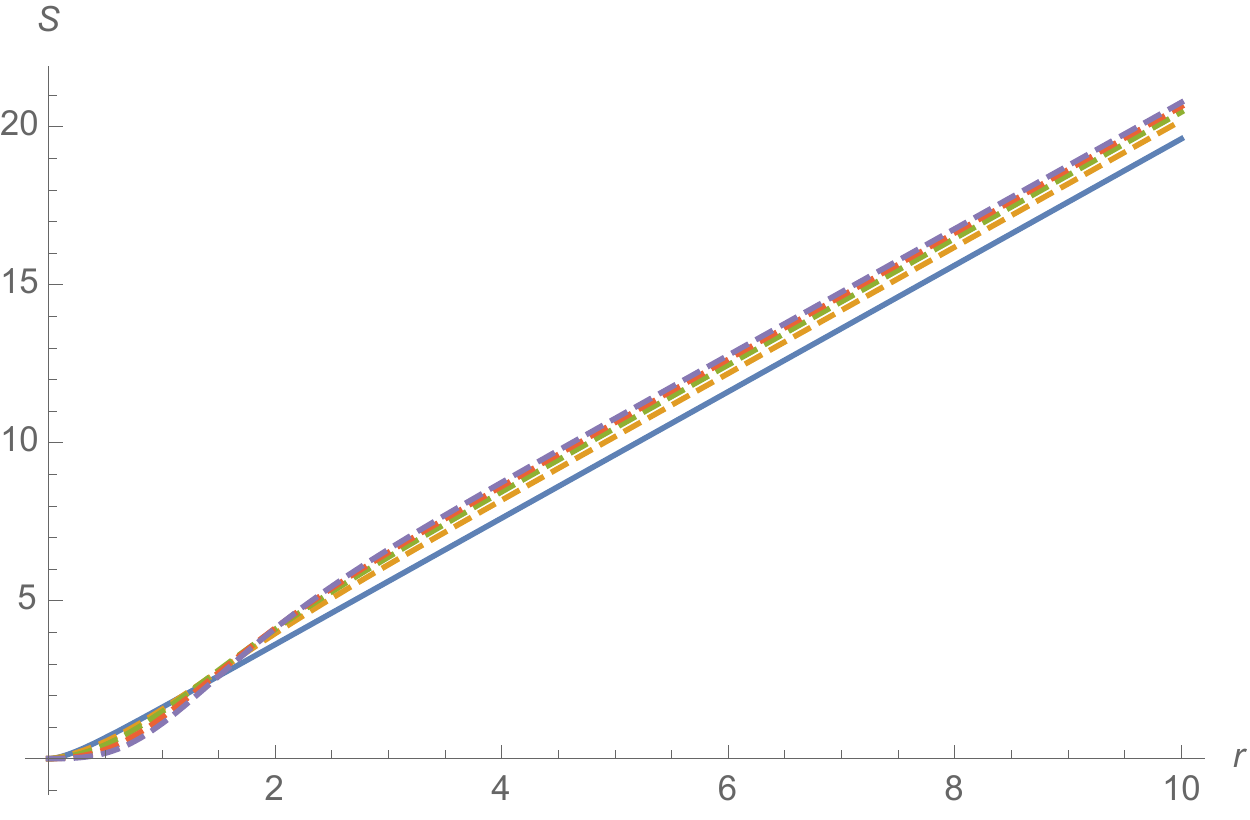}
    \end{subfigure}
\begin{subfigure}
\centering
    \includegraphics[width=0.45\textwidth]{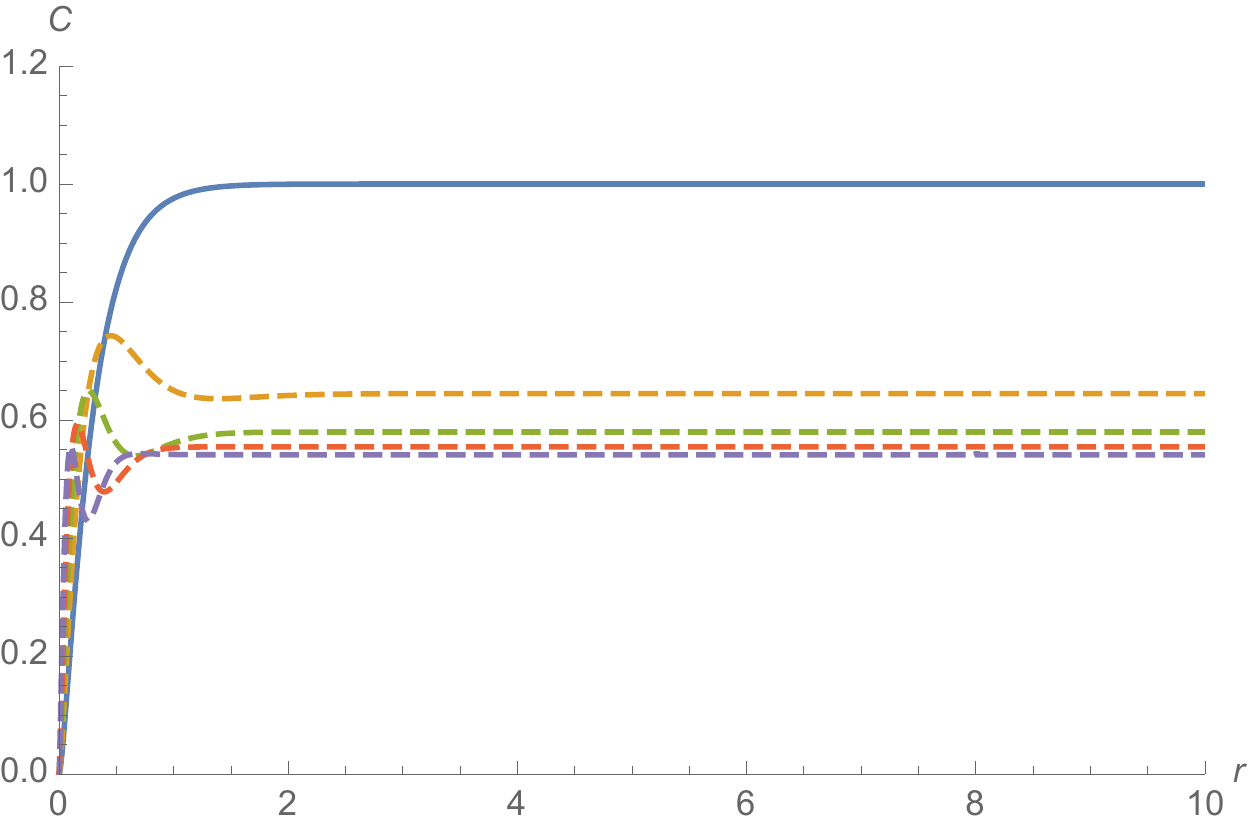}
\end{subfigure}    
 \caption{Left: Von Neumann entropy as function of the parameter $r$ for $n_0=0,1,2,3,4$. Once again we observe a universal linear trend for late times, consistent with our expectations. Right: Capacity of entanglement as a function of the parameter $r$ for $n_0=0,1,2,3,4$. The late time behavior is again universal but its quantitative precision is unclear. Also notice that for early times we observe the development of local minima and maxima, which we comment on further in the main text.}
 \label{sl2 capa}
\end{figure}

Within the regime of the aforementioned approximation, one can compute the associated von Neumann entropy and capacity of entanglement. We begin by computing the Rényi entropies as shown below 
\be \label{renyi sl2}
S^{(q)}=\frac{1}{1-q}\ln{\sum_n \abs{\varphi_n}^{2q}}\simeq\frac{1}{1-q}\ln{\left(\frac{\text{Li}_{-qn_0}(\tanh^{2q}r)}{(\cosh{r})^{2q(1+n_0)}(n_0\Gamma(n_0))^{q}}\right)}~.
\ee
It is then straightforward to obtain the von Neumann entropy and capacity of entanglement, although the resulting expressions are quite lengthy and for that reason their presentation here is omitted. However, we can still generate graphical data from them which are of particular value and interest. These are shown in figure \ref{sl2 capa}.
An interesting observation is that the capacity of entanglement exhibits some non-universal features at early times. In particular, we find that as $n_0$ grows the plots develop local minima and maxima, for which a physical explanation does not appear to be straightforward. Even though the quantitative features of these plots are approximate, we have confidence in the qualitative ones even for early times. One can easily confirm this claim by employing an approach that is in some sense the reverse of our approximation and namely by numerically performing the sum in \ref{renyi sl2}
up to some finite value of $n$. By doing so we can capture the correct qualitative behavior for early times as justified by figure \ref{r-dependence} and indeed verify the existence of these mysterious local extrema. The entanglement entropy also exhibits some non-universal features at early times albeit much more subtle and much less puzzling. 

Another important comment regarding the capacity of entanglement has to do with the differences in the saturation value for different $n_0$. Even though we do not necessarily expect these values to be accurate, it is interesting to consider whether their differences are a feature of the various SU(1,1) representations rather than a byproduct of our approximation. It would also be interesting to consider whether this behavior can be captured in more physical terms, for example by the quasi-particle picture provided in \cite{Nandy:2021hmk}.

\subsection{SU(2)}

For the case of SU(2) we will simply repeat the process we illustrated above using the appropriate coherent states which we presented in section 3. Note that, unlike the SU(1,1) case, the sums we will have to perform are always finite, because the discrete series representations of SU(2) have a finite number of Fock states. This makes things significantly more simple, as there will be no approximations required, but rather an exact result can be obtained for any SU(2) representation.

\begin{figure}[ht]
    \centering
    \includegraphics[scale=0.6]{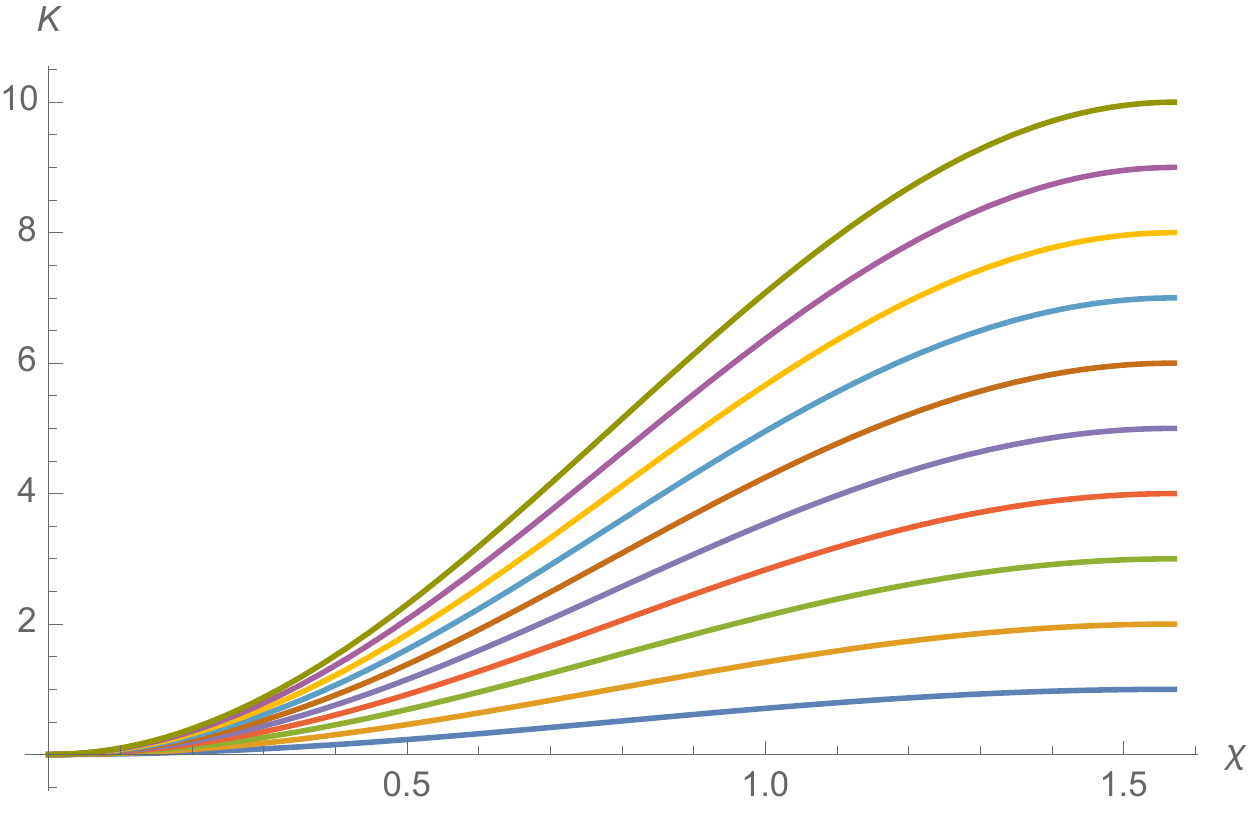}
    \caption{Krylov complexity as a function of $\chi$ for $\frac{1}{2}\leq j\leq 5$. We observe that in all case the Krylov Complexity grows to its maximum value at $\chi=\frac{\pi}{2}$ as expected.}
    \label{krylov su2}
\end{figure}

Once again we start by focusing on Krylov complexity, for which by definition we have
\be
K=\sum_{N_2=0}^{2j}N_2 \abs{F_{N_2}}^2=2j\sin^2{\chi}~.
\ee
The Krylov complexity is shown for different values of $j$ in figure \ref{krylov su2}.

\begin{figure}[ht]
\begin{subfigure}
 \centering
    \includegraphics[width=0.45\textwidth]{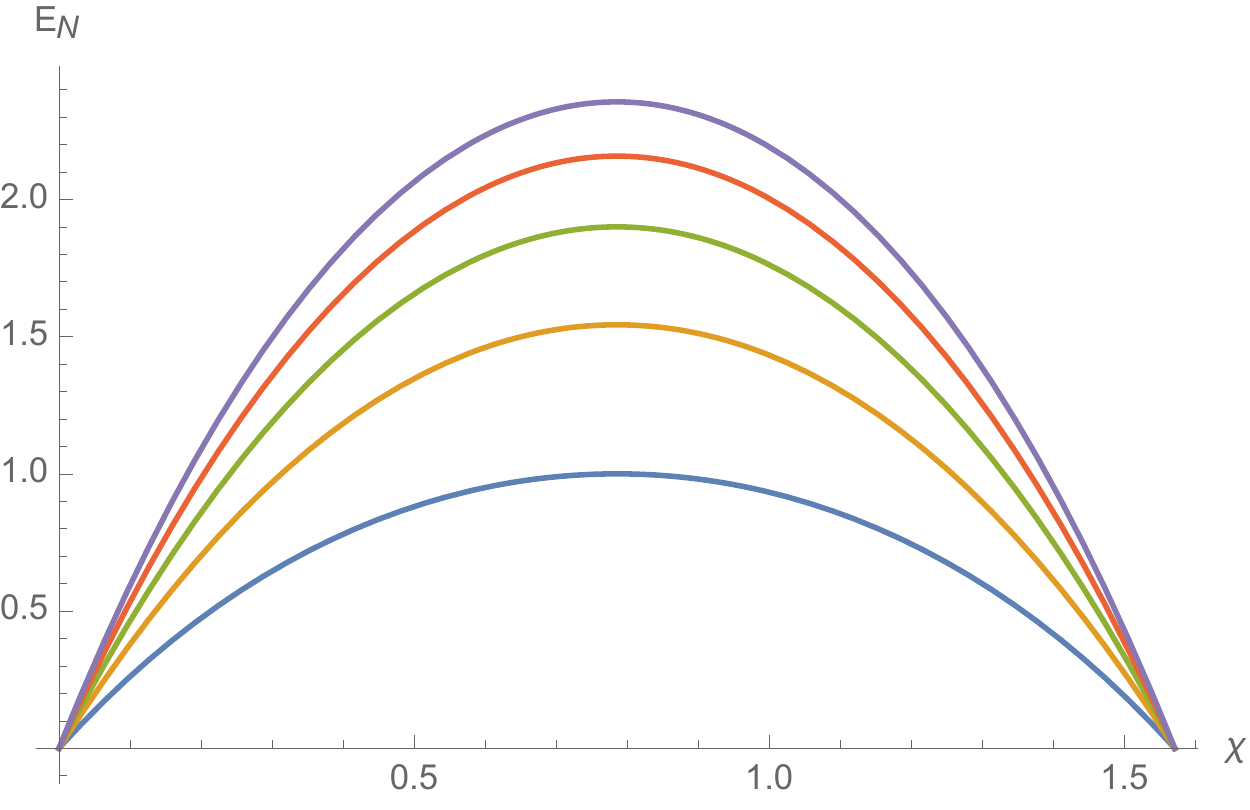}
\end{subfigure}
\begin{subfigure}
\centering
  \includegraphics[width=0.45\textwidth]{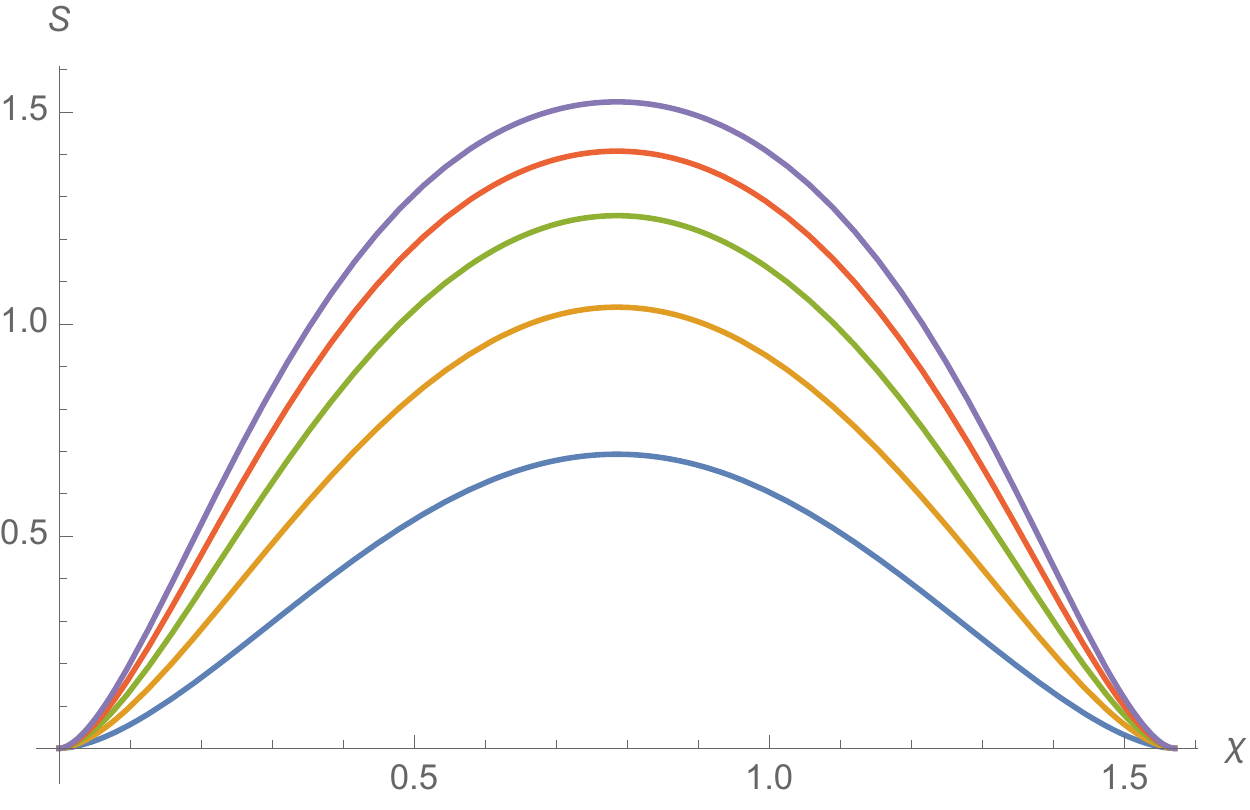}
\end{subfigure}
 \caption{Left: Logarithmic negativity as a function of $\chi$ for $\frac{1}{2}\leq j \leq \frac{5}{2}$. Right: Entanglement entropy as a function of the parameter $\chi$  for $\frac{1}{2}\leq j \leq  \frac{5}{2}$. }
\label{SU(2)_negativity}   
\end{figure}

\begin{figure}[b!]
    \centering
    \includegraphics[scale=0.6]{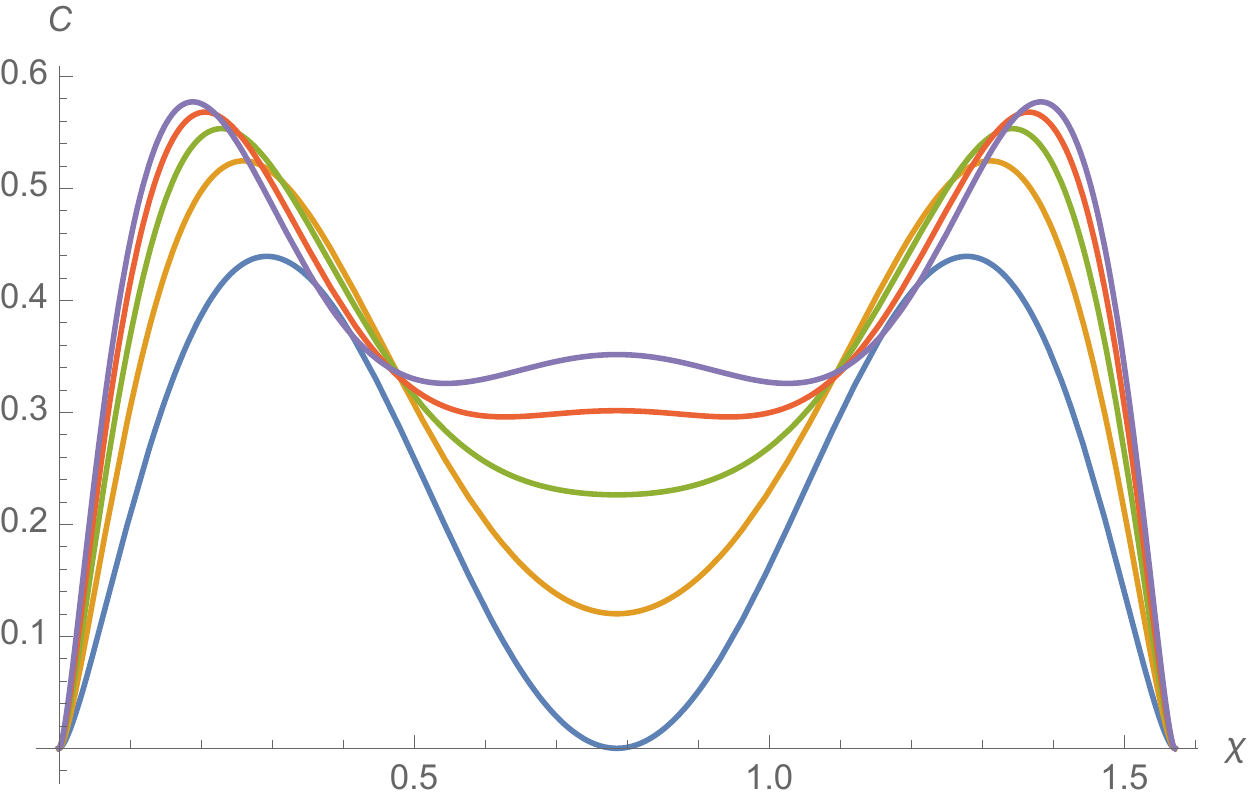}
    \caption{Capacity of entanglement as a function of the parameter $\chi$  for $\frac{1}{2}\leq j \leq \frac{5}{2}$ }
    \label{SU(2)_capacity}
\end{figure}

 Proceeding to the computation of negativity, following our results for SU(1,1), we simply have to perform the sum
\be
E_\mathcal{N}=2\log_2\sum^{2j}_{N_2=0} |F_{N_2}|~.
\ee
For general $j$ the sum does not assume a closed form, but since it is always finite one can easily obtain an answer for any given value of $j$. Below we present the results for a few of them in figure \ref{SU(2)_negativity}.
The entanglement entropy and capacity of entanglement can be straightforwardly obtained from the corresponding definitions and in figure \ref{SU(2)_capacity} we present the resulting plots for different values of $j$.

 Similarly to the case of SU(1,1) it is apparent that for all the quantum information tools under study there are certain universal features that characterize them. The important distinction with the SU(1,1) example is that we observe a finite instead of an infinite growth. This difference originates from the fact that for SU(2) there is a finite number of Fock states that we are using to describe this process. Having explained the identification between the coherent states and Krylov basis it follows that the operator only has a ``limited space" in which it can grow. This is clearly reflected in the above figures, in support of this picture. An important comment on our results concerns the capacity of entanglement which, similarly to the SU(1,1) case, exhibits some sensitivity to representation specific information. Namely, it is evident that for the representations with $j=\frac{1}{2},1,\frac{3}{2}$ there is a minimum at $\chi=\frac{\pi}{4}$, whereas for representations with $j\geq2$ this turns to a local maximum. The question of whether there is some physical picture that can explain this behavior is also relevant here and once again an answer could potentially be provided in terms of quasi-particle entanglement between EPR pairs \cite{Nandy:2021hmk}.

\section{Conclusions and discussion}
We have considered the computation of several quantum information tools for systems with Lie symmetry. More specifically, we have shown that for systems with SU(1,1) and SU(2) symmetry we can easily obtain the associated Krylov complexity, negativity, entanglement entropy and capacity of entanglement for any discrete series representation by studying the properties of the corresponding two-mode coherent states. Furthermore, we illustrated that all of these quantities exhibit certain universal features in agreement with the universal operator growth hypothesis. For the SU(1,1) case we used an approximation which we argued is accurate for late times. Despite that, we were also able to obtain some qualitative features of the early time behavior of such systems which would be interesting to explain from a physical perspective. Our approach was similar in the case of SU(2) for which there was no need for an approximation and hence all results are exact. Once again we concluded that apart from the universal aspects of the growth that we can observe, there are certain features that differ between the various representations. Namely, the capacity of entanglement seems to be sensitive to this kind of information and in fact exhibits a rather interesting transition for $j=2$. To clarify, these observations are still consistent with the operator growth hypothesis, as the operator of any system in a given representation of the groups we considered will behave in the same way. However, it appears that there are certain traits in the quantities we studied that would allow one to discern which particular representation the system is in, which in turn implies that this aspect of operator growth is not universal.

There are several directions that are worthwhile to investigate in the future. In particular, here we have considered only two cases of Lie groups, so it would be interesting to determine whether this approach works more generally. However, not all Lie groups admit a two-mode coherent state representation, as is the case for the Heisenberg-Weyl group for example. One could still assert that the quantities we defined using the two-mode states would be valid in such cases, however, a more detailed analysis is required to prove this claim rigorously. Making progress in that direction would be an important step forward, as there are already works that have considered Krylov complexity in other setups using the Lanczos algorithm. For example in \cite{Dymarsky:2021bjq} the authors consider different models of CFTs, such as 2d CFTs, free field and holographic models. We know the symmetry groups of these theories, so in principle it is possible to use our approach in order to compare the results for Krylov complexity and possibly complement the picture by computing the other quantum information tools that we have discussed. Recently there have also been works on Nielsen complexity that are similar in spirit in their use of coherent states or the symmetry of the system in general \cite{Guo:2018kzl,Caputa:2018kdj,Chagnet:2021uvi,Koch:2021tvp,Basteiro:2021ene}. It would be interesting to further explore the connection between these two approaches to complexity given their similarities and determine the point at which they diverge and what this can teach us about the field as a whole.

 As we have stressed in previous sections, here we have only considered the discrete series representations of the groups under study. So, another question that naturally arises from our tools from symmetry prescription is whether it can be extended to include other representations as well. This would require constructing the coherent states for these representations and using them to define the quantum information quantities we are interested in. It is already well known how to construct the states (as reviewed in \cite{Perelomov:1986tf} for example), although the process is more involved compared to the discrete series case, consequently leading to several subtleties.
 
In light of some recent advances in the field of dS/CFT \cite{Strominger:2001pn,Witten:2001kn} we would like to point to the potential relevance of our results for SU(2) in this direction. In particular, the authors of \cite{Hikida:2021ese} advocate that the CFT dual of dS$_3$ is given by an SU(2) Wess-Zumino-Witten model in the large central charge limit. As such, it is very intriguing to consider whether our approach provides a natural candidate for exploring holographic complexity in dS/CFT.

 Finally, as shown in \cite{Caputa:2021sib}, Krylov complexity can be interpreted as a volume on the space of coherent states. Therefore, it would be interesting to consider whether the other quantum information tools that we have discussed also admit an interpretation in terms of the coherent state geometry. Given that we are using the coherent state properties to compute these quantities, it appears intuitively plausible that they should indeed possess some geometric interpretation. However, at the time of writing it is not clear how these identifications can be performed in a rigorous manner.

\section*{Acknowledgements}
The author wishes to thank Aditya Bawane, Jan Boruch, Dongsheng Ge and Javier Magan for helpful discussions and comments. Special thanks to Paweł Caputa for his continuous guidance through all the stages of this work. DP is supported by NAWA ``Polish Returns 2019".

\appendix 

\section{CFT construction of SU(1,1) cohererent states}\label{appendix}
This appendix is devoted to an alternative construction of the coherent states presented in section 2 using CFT techniques. This is feasible as the global symmetry group of 2d CFTs is SU(1,1) and hence the symmetry arguments that were presented previously are implicitly encoded in the CFT formalism that we use below. For a similar construction of the coherent states associated with the full Virasoro symmetry of 2d CFTs see \cite{Caputa:2021ori}. Our starting point is a state of highest weight $h$, $\ket{h}=\Op_{-h}\ket{0}$, where  $\Op_{-h}$ is a mode of a chiral primary operator with dimension $h$. This means that there is an expansion of the form 
\be
\Op(z)\ket{0}=\sum_{n=0}^\infty z^n \Op_{-h-n}\ket{0}~,
\ee
which we will make use of. Furthermore, we know that the modes have the following properties
\be
[L_{m},\Op_{-h}]=(h(m-1)-m)\Op_{-h+m}, \quad m=\{-1,0,1\}~,
\ee
\be
[\Op_m^{(i)},\Op_n^{(j)}]=\delta_{m+n,0}d^{ij}\binom{m+h-1}{2h-1} +\sum_k C^{ij}_k p^{ij}_k(m,n)\Op^{k}_{m+n}~,
\ee
where we have adopted the usual notation for the SU(1,1) generators in the CFT language in terms of $L_m$ and $d^{ij}$ are structure constants that can be set to $\delta^{ij}$.

We are interested in the action of the displacement operator on these states, which as we argued previously is given by
\be
D(\xi)= e^{\xi L_{-1}-\Bar{\xi}L_1}~.
\ee
Using the BCH formula, the action of the displacement operator on the states can be written as 
\be
e^{zL_{-1}}e^{aL_0}e^{\Bar{z}L_1}\ket{h} =e^{zL_{-1}}e^{aL_0}e^{\Bar{z}L_1}\Op_{-h}\ket{0}~,
\ee
where $a,z$ are numbers to be determined. It is easy to verify that this can be reduced to 
\be
e^{ah}e^{zL_{-1}}\Op_{-h}\ket{0}~,
\ee
as commuting $L_1$ acting on $\ket{h}$ yields zero and $L_0\ket{h}=h\ket{h}$. Thus, by inserting the identity we obtain
\begin{align}
&e^{ah}e^{zL_{-1}}\Op_{-h}e^{-zL_{-1}}e^{zL_{-1}}\ket{0}=\\
&e^{ah}\sum_{n=0}^\infty z^n\Op_{-h-n}\ket{0}=e^{ah}\Op(z)\ket{0}=\ket{z,h}~.
\end{align}
The factor $e^{ah}$ can be simply regarded as the normalization of the state and can be obtained by imposing the condition
\be \label{normalization}
\bra{0}\Op^\dagger(z)e^{2ah}\Op(z)\ket{0}=1 \Rightarrow e^{2ah}=\frac{1}{\braket{\Op^\dagger(z)\Op(z)}}~.
\ee
In order to compute the correlator we will make use of the relation 
\be
\Op^\dagger(z)=\Bar{z}^{-2h}\Op(1/\Bar{z})
\ee
and the standard result for two-point functions in 2d CFT 
\be
\braket{\Op(z)\Op(w)}=\frac{1}{(z-w)^{2h}}~,
\ee
which then imply
\begin{equation}
    \Bar{z}^{-2h}\braket{\Op(1/\Bar{z})\Op(z)}=\Bar{z}^{-2h}(\frac{1}{\Bar{z}}-z)^{-2h}=(1-\abs{z}^2)^{-2h}~.
\end{equation}
Substituting this result in (\ref{normalization}) we obtain
\be
e^{ah}=\frac{1}{\sqrt{\braket{\Op^\dagger(z)\Op(z)}}}=(1-\abs{z}^2)^h~.
\ee
Let us now consider the coefficients given by the projection of the states $\ket{z,h}$ on the modes of $\Op(z)$
\be \label{cft coef}
\psi_n=\bra{0}\Op^\dagger_{-h-n}\ket{z,h}=\frac{\braket{\Op_{h+n}\Op(z)}}{\sqrt{\braket{\Op^\dagger(z)\Op(z)}}}~.
\ee
The numerator on the RHS is 
\begin{align}
    \braket{\Op_{h+n}\Op(z)}&=\braket{[\Op_{h+n},\Op(z)]}+\braket{\Op(z)\Op_{n+h}}\\
    &=\braket{[\Op_{h+n},\sum_m z^m\Op_{-h-m}]}= z^n\binom{2h+n-1}{2h-1}~,
\end{align}
and by the appropriate substitutions we can rewrite (\ref{cft coef}) as
\be
\psi_n=(1-\abs{z}^2)^{h}z^n\binom{2h+n-1}{2h-1}~. 
\ee
Using the identifications $z=e^{i\phi}\tanh{r}$ and $2h-1=n_0$ these precisely match the coefficients we computed in section 2, with $\psi_n=e^{in\phi} \varphi_n$. 

In summary, one can construct the same set of coherent states that we derived in section 2 by thinking exclusively in terms of 2d CFT quantities. In particular, the coherent states themselves can be obtained by the action of a chiral primary operator on the vacuum state and the coefficients of their decomposition in terms of modes are given by correlators whose form is fixed by the conformal symmetry.

\bibliographystyle{JHEP}
\bibliography{references.bib, misc.bib}
\end{document}